\documentclass[printer]{aa} 

\usepackage{natbib}
\usepackage{graphicx}
\usepackage{txfonts}
 \title{High-resolution spectropolarimetric observations of the temporal evolution of magnetic fields in photospheric bright points}
	\authorrunning{P. H. Keys et al.}
	\titlerunning{Rapid amplification of magnetic fields in bright points}
   \author{P.~H.~Keys\inst{1}, A.~Reid\inst{1}, M.~Mathioudakis\inst{1}, S.~Shelyag\inst{2}, V.~M.~J.~Henriques\inst{3,}\inst{4,}\inst{1}, R.~L.~Hewitt\inst{1}, D.~Del~Moro\inst{5}, S.~Jafarzadeh\inst{3,}\inst{4}, D.~B.~Jess\inst{1,}\inst{6}, \and M.~Stangalini\inst{7,}\inst{8}\\
         }

\institute{Astrophysics Research Centre, School of Mathematics and Physics, Queen's University Belfast, Belfast, BT7 1NN, Northern Ireland, U.K.
\and
School of Information Technology, Faculty of Science, Engineering and Built Environment, Deakin University, 221 Burwood Highway, Burwood, VIC 3125, Melbourne, Australia
\and
Rosseland Centre for Solar Physics, University of Oslo, P.O. Box 1029 Blindern, NO-0315 Oslo, Norway
\and
Institute of Theoretical Astrophysics, University of Oslo, P.O. Box 1029 Blindern, NO-0315 Oslo, Norway
\and
Dipartimento di Fisica, Universit\`a degli Studi di Roma ``Tor Vergata'', via della Ricerca Scientifica 1, 00133 Roma, Italy
\and
Department of Physics and Astronomy, California State University Northridge, Northridge, CA 91330, U.S.A
\and
INAF-OAR National Institute for Astrophysics, Via Frascati 33, 00078 Monte Porzio Catone (RM), Italy
\and
ASI Agenzia Spaziale Italiana, Via del Politecnico snc, I-00133 Rome, Italy
}

\offprints{P. Keys, \email{p.keys@qub.ac.uk}}
   \date{Received August 2019; Accepted November 2019}

\begin{document} 
  \abstract
   {Magnetic bright points (MBPs) are dynamic, small-scale magnetic elements often found with field strengths of the order of a kilogauss within intergranular lanes in the photosphere. 
}
   {Here we study the evolution of various physical properties inferred from inverting high-resolution full Stokes spectropolarimetry data obtained from ground-based observations of the quiet Sun at disc centre. 
}
   {Using automated feature-tracking algorithms, we studied 300 MBPs and analysed their temporal evolution as they evolved to kilogauss field strengths. These properties were inferred using both the NICOLE and SIR Stokes inversion codes. We employ similar techniques to study radiative magnetohydrodynamical simulations for comparison with our observations.
}
   {Evidence was found for fast ($\sim$30~--~100s) amplification of magnetic field strength (by a factor of 2 on average) in MBPs during their evolution in our observations. Similar evidence for the amplification of fields is seen in our simulated data.
}
   {Several reasons for the amplifications were established, namely, strong downflows preceding the amplification (convective collapse), compression due to granular expansion and mergers with neighbouring MBPs. Similar amplification of the fields and interpretations were found in our simulations, as well as amplification due to vorticity. Such a fast amplification will have implications for a wide array of topics related to small-scale fields in the lower atmosphere, particularly with regard to propagating wave phenomena in MBPs.

}


   \maketitle

				
\section{Introduction}
\label{Intro}
Magnetic bright points (MBPs) are ubiquitous in the quiet solar photosphere \citep{Solanki1993}. Theory suggests that convection can lead to kilogauss fields in these small-scale features by a process termed `convective collapse' \citep{Spruit79}. The basis of this process is that flux within intergranular lanes is subject to strong downflows, which results in the flux tube reducing in size to balance external forces from surrounding material on the tube. Given that the flux tube is partially evacuated and there is heating from the surrounding hot granular walls, this results in the flux tube appearing as a localised intensity enhancement within the intergranular lane. Interestingly, recent work on magnetohydrodynamic (MHD) simulations \citep{Calvo2016} indicates that it is possible to have localised intensity enhancements within the photosphere in the absence of magnetic fields (non-magnetic BPs). These bright points are caused by a reduced mass density within a swirling downdraft funnel and are at a scale of 60~--~80~km. The complexity of possible intensity enhancements highlights the need for continued study of the formation processes of these features in the photosphere.

Due to their small spatial scales \citep[diameters of $\sim$100~--~300\,km;][]{SanchezAlmeida2004, Utz2009, Crockett2010}, MBPs were not studied rigorously until adaptive optics and image restoration techniques became mainstream. Observational evidence for convective collapse emerged several years after the initial theory was proposed \citep{Solanki1996, BellotRubio2001, Nagata2008}. Using {\textit{Hinode}} data of a single MBP, \citet{Nagata2008} searched for evidence of line-of-sight (LOS) velocity increases just prior to an intensity enhancement in G-band images of a region with a magnetic flux concentration. In doing so, the authors were able to show the process of convective collapse within this MBP. Subsequent work \citep{Fischer2009}, used the same approach and found that the radii of 49 MBPs reduced on average from 0$\arcsec$.43 to 0$\arcsec$.35 which lead to field strengths of up to 1.65\,kG. Similarly, \citet{Narayan2011} found that targeting small-scale transient downflows as a selection criteria in a plage region would identify short-lived MBPs. A more extended study used near seeing free {\textit{SUNRISE/IMAX}} data \citep{Utz2014} to examine the general properties of the magnetic field and LOS velocity evolution for 200 MBPs. They found that MBPs are formed when the LOS velocity increases to an average of 2.4~km\,s$^{-1}$, although only 30\% of MBPs reach kilogauss field strengths and show downflows of over 1~km\,s$^{-1}$ over 46\% of their lifetimes.

There are a few methods by which MBPs are observed and theorised to disintegrate. One suggestion is that the MBPs split apart, weakening the field to a point where it falls below the equipartition field strength. As a result, they will no longer appear as an intensity enhancement, effectively disintegrating the MBP through turbulent convection. Likewise, it has been suggested that in a similar manner to turbulent convection splitting the MBP, flux may be removed from the MBP through diffusion. Thin tubes could also disintegrate through the interchange instability, which can be overcome through vortical flows near the flux tube \citep{Parker1975, Schussler1984}. Increasing evidence from simulations \citep{Shelyag2011} and observations \citep{Requerey2017, Requerey2018} shows that vortical flows are present in and/or in the vicinity of MBPs and that a weakening of the vortex flow can lead to the weakening of the magnetic element prior to fragmentation.

Another possibility is that small-scale reconnection events, such as Ellerman Bombs \citep {Nelson2013, Reid2016}  and other cancellation events \citep{Sheminova2000, Borrero2010} remove flux from the photosphere releasing energy to the surroundings. The MBP disintegration process can be caused by a reversal of the convective collapse process \citep{GrossmannDoerth1998, Steiner1998, Takeuchi1999, BellotRubio2001}, whereby the strong downflow within the flux tube rebounds in the deep photosphere causing an upwardly moving shock within the flux tube. This results in an upflow within the flux tube, weakening the field until it falls below its  equipartition value. \cite{Utz2014} state that only 16\% of MBPs have upflows prior to disintegration. It is probable that all of these processes have varying degrees of significance in the disintegration process of MBPs, although more work needs to be carried out to determine the relative importance of each.
 
MBPs are highly dynamic, with transverse velocities of around 1~km\,s$^{-1}$ \citep{Utz2009, Keys2011}. It is estimated that there are approximately 0.97~MBPs per Mm$^{2}$ covering $\sim$0.9\%\,--\,2.2\% of the solar surface \citep{SanchezAlmeida2010}. The lifetimes of MBPs range from 90~s to 3~minutes dependent on the cadence of the instrument used for the observations, the technique to track them and the environment in which they are found \citep{Berger96, Utz2010, Keys2011, Keys2014}. 

There have also been a few studies that investigate the magnetic field of MBPs \citep[e.g. ][]{BellotRubio2001, Utz2013, Requerey2014}. Statistical analysis of the magnetic field strength  of MBPs  find a bimodal distribution \citep{Utz2013, Keys2019}, with a group of `weak' field MBPs peaking at approximately 300~--~600G and a `strong' field group at 1100~--~1300G. This bimodal distribution is often attributed to the process of convective collapse, with the weaker group being MBPs that have yet to go through convective collapse, or going through the reverse process towards disintegration, while the strong group are MBPs that have gone through collapse to achieve kilogauss field strengths. A recent work by \citet{Keys2019} found a bimodal distribution of magnetic fields in observations, however, they did not find a similar distribution in MURaM radiative MHD simulations of a domain with 200~G and 50~G initial fields. The authors attributed this to a combination of flux emergence and diffusion of the emerging flux within the observations, in that the weak group results from `new' flux emerging in the field-of-view (FOV), dispersing before it can amplify to kilogauss field strengths. The study suggests that convective collapse was not solely responsible for the observed bimodal distributions of magnetic fields in MBPs, and that the diffusion of emerging flux plays a role in MBP evolution. 

It should be noted that distributions aside from the bimodal one have been observed for MBPs as well. Results from \citet{Beck2007} of MBPs in the moat of a sunspot find a flat distribution with field strengths in the range 500~--~1400~G. \citet{Viticchie2010} find a single distribution while observing a quiet Sun region with the Interferometric BIdimensional Spectrometer \citep[IBIS][]{Cavallini2006}. Similar to \citet{Keys2019}, a study of 3D MHD simulations with the Copenhagen stagger code \citep{GalsgaardNordlund1996} by \citet{Criscuoli2014} find a single distribution of magnetic field strengths in MBPs. The authors find that the shape of the distribution changes when accounting for spatial degradation and misalignment in data. Therefore, there are a range of factors that should be considered in the case of the appearance of magnetic field distributions in MBPs.

Their magnetic and dynamic nature indicates that MBPs are the source of MHD waves \citep[][to name a few]{MartinezGonzalez2011, Jess2012, Stangalini2013a, Stangalini2013b, Mumford2015, Stangalini2015, Jafarzadeh2017a}. They have been observed to have frequent excursions above 3~km\,s$^{-1}$, in super diffusive L{\'{e}}vy-flights \citep{Chitta2012, Giannattasio2013, Keys2014}, with recent work \citep{Jafarzadeh2017b} suggesting that the diffusion of small-scale magnetic features in the photosphere is related to the region they are observed (i.e. within network cells and close to regions of flux emergence). Velocity excursions above 3~km\,s$^{-1}$ are sufficient to generate kink modes \citep{Choudhuri93}. Indeed, kink modes have been observed and studied in MBPs \citep{Stangalini2013a, Stangalini2013b, Stangalini2015} with studies finding clear evidence for upward propagation to the chromosphere with frequencies above 2.6~mHz. Sausage modes in MBPs are more difficult to observe, in part due to the small size of MBPs and the fact that the fractional variations in area that are synonymous with sausage modes are of the order of only a few percent \citep[as observed in pores][]{Dorotovic2008, Grant2015, Freij2016, Keys2018}. Nevertheless, there are some studies which may have observed signatures of sausage modes in MBPs \citep{Requerey2014, Jafarzadeh2017a}. Finally, the first observational evidence for Alfv{\'{e}}n waves in the lower solar atmosphere \citep{Jess2009} was observed in an MBP group by looking for asymmetries in the the FWHM of a spectral line scan. The energy estimate for this Alfv{\'{e}}n wave was found to be 15000~W\,m$^{-2}$, thus, their ubiquity means that MBPs could be a significant source of heating in the solar atmosphere.

In this work, we study the evolution of specific MBP properties in detail, by focusing on several interesting examples we find in our dataset. Our observational findings were then compared to results from simulated datasets. In Section~\ref{ObsSims}, we describe our observations and simulations, while the methodology we apply is described in Section~\ref{Methods}. We describe the key results and their implications in Section~\ref{Results} before our concluding remarks in Section~\ref{conc}.

				
\section{Observations and simulations}
\label{ObsSims}
Data of a quiet Sun region at disc centre was acquired with the 1~m Swedish Solar Telescope \citep[SST;][]{Scharmer2003} on 2014 July 27$^{th}$ from 14:18\,UT until 15:11\,UT with an initial pointing of N0.14, W4.5. The CRisp Imaging SpectroPolarimeter \citep[CRISP;][]{Scharmer2006, Scharmer2008} was used to sample the Fe~{\sc{i}} 6301~{\AA} and 6302~{\AA} line pair in full Stokes spectropolarimetry mode at 32 wavelength positions with a spectral FWHM of 53.5~m{\AA}. For the 6301~{\AA} line, the observed wavelengths in m{\AA} were: $-$462, $-$385, $-$308, $-$231, $\pm$154, $\pm$115, $\pm$77, $\pm$38, 0, 192, 269, 346, from the core position. A marginally smaller step size was used for the 6302~{\AA} line to account for the slightly narrower line leading to sampling locations at $-$333, $\pm$259, $\pm$185, $\pm$111, $\pm$74, $\pm$37, 0, 148, 222, 296, 407 from the core (also in m{\AA}). Good continuum points, avoidance of smaller blends, sampling the telluric line in the red wing of 6302, and keeping a common denominator for the step size per line comprised the selection criteria for these wavelength steps. An unusually high number of exposures of nine per wavelength, targeting signal-to-noise, were used. A total of 92 complete full Stokes scans were taken over the duration of the observations. 

Image reconstruction was performed using the Multi-Object Multi-Frame Blind Deconvolution technique \citep{Lofdahl2002,vanNoort2005} within the CRISPRED data reduction pipeline \citep{delaCruzRodriguez2015} which includes de-rotation, destretching \citep[as in][]{Shine1994}, and the extended reconstruction scheme of \citet{Henriques2012}, with the latter being especially important in preventing seeing induced cross-talk from polarimetric data. The demodulation procedure included a current time-varying telescope model obtained for the observed wavelength range, as that described in \citet{Selbing2010} and \citet{Schnerr2011}, and a calibration for the table optics acquired the same day. The post-reduction cadence for the scans is around 33~s. The effective FOV of the data was approximately 50$'' \times$50$''$ with a spatial sampling of 0.$''$059 pixel$^{-1}$. Additionally, 1~s cadence photospheric images were acquired using a Ca wide band filter centred on 395.37 nm (FWHM 1.0 nm). The same dataset is employed in the study by \citet{Keys2019}, where a bimodal distribution in the magnetic field strengths were observed in the tracked MBPs.

\begin{figure*}[h!]
\makebox[\linewidth]{
   \includegraphics[width=0.9\linewidth]{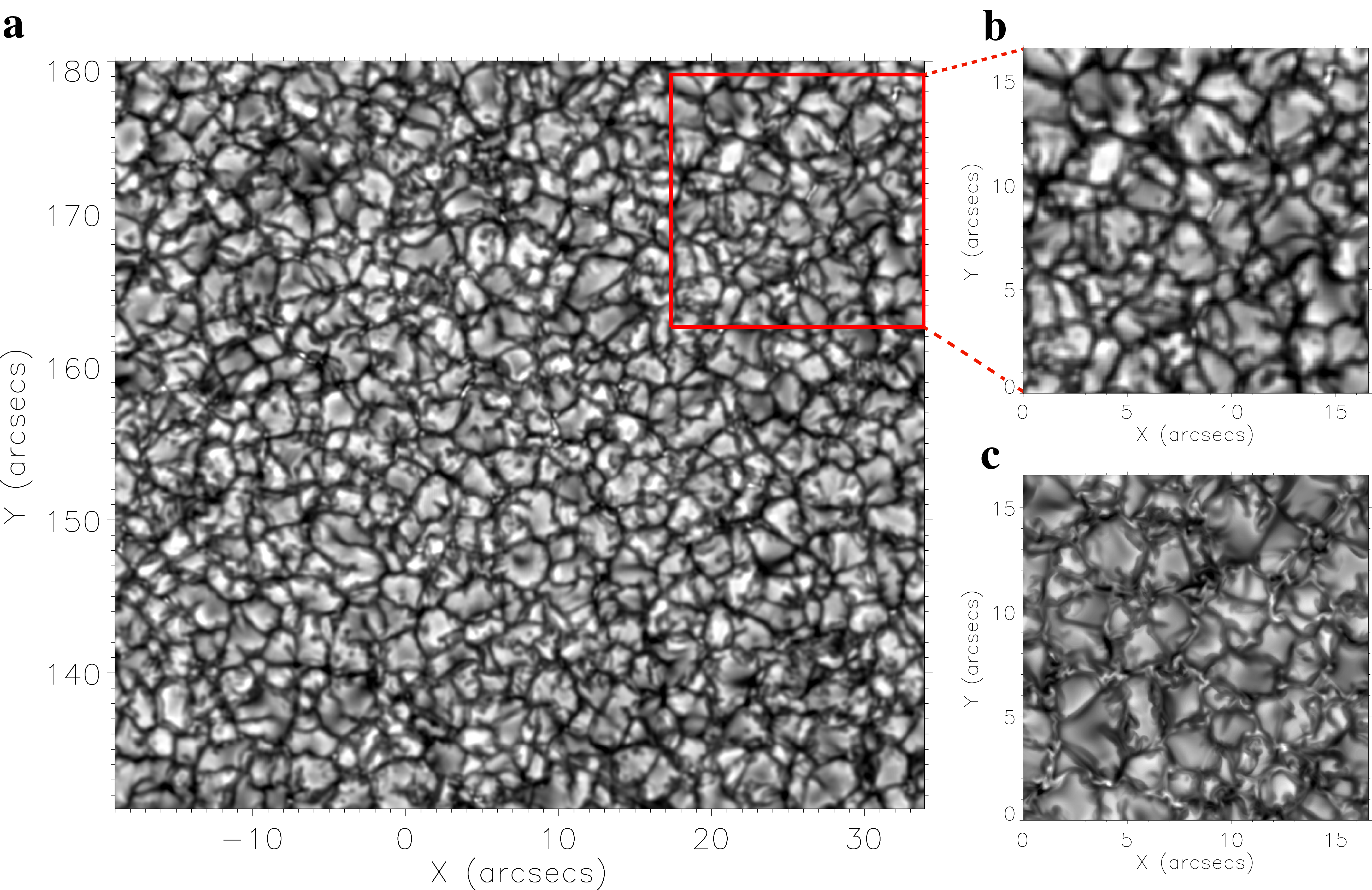}
	}
     \caption{Panel \textbf{a} shows the full field-of-view of our data set as observed with CRISP at the SST on 2014 July 27$^{th}$ from around 14:18\,UT until 15:11\,UT in the 6301{\AA} and 6302{\AA} line pair with this image taken at 6301.0392~{\AA} in the line scan. The target was the quiet Sun at disc centre. The red box in panel \textbf{a} shows the region used for the cut out displayed in panel \textbf{b}. Panel \textbf{b} is a 16$\arcsec \times$ 16$\arcsec$ region of the field-of-view. This cut out is shown as a direct comparison of the MURaM simulations presented in panel \textbf{c}, which have the same 16$\arcsec \times$ 16$\arcsec$ spatial dimensions. These MURaM simulations have an initial 200~G field, and a spatial resolution of 25~km\,pixel{$^{-1}$}.
}
     \label{Fig1}
\end{figure*}

The MURaM radiative MHD code \citep{Vogler2005} was used to produce our simulated data. This code solves large-eddy radiative three-dimensional MHD equations on a Cartesian grid using a fourth-order Runge-Kutta scheme to advance the numerical solution in time. The numerical domain is resolved by 480$\times$480$\times$100 grid cells representing a physical size of 12$\times$12~Mm$^2$ in the horizontal direction, 1.4~Mm in the vertical direction, respectively. The starting point for our simulations is a well-developed non-magnetic ($B=0$) snapshot of photospheric convection taken approximately 2000~s (about 8 convective turnover timescales) from the initial plane-parallel model. At this stage, a uniform magnetic field of 200~G before a sequence of 339 snapshots was recorded. Each snapshot was separated in time by $\sim$17~s. The final simulations sequence covers approximately 90~minutes of physical time, corresponding to $\sim$10~--~20 granular lifetimes. The NICOLE \citep{SocasNavarro2015} code is employed in synthesis mode to produce the synthetic Stokes I, Q, U and V profiles for the models obtained from MURaM for the 6301~{\AA} and 6302~{\AA} line pair with the same wavelength step as our CRISP observations. Figure~\ref{Fig1} shows a sample image from both our observations and simulated datasets.

\section{Methods}
\label{Methods}
To isolate the MBPs, a threshold tracking algorithm was employed \citep[see][for details of this code]{Crockett2010} on both the Fe~{\sc{i}} Stokes I images and the Ca wide band data. At the most basic, the code isolated bright features within the image, then performs an intensity threshold (at 5$^\circ$ intervals) across the isolated features to determine whether the feature is an MBP or not. The intensity profile of an MBP will be different to that of, say, a granule, as the MBP will have a sharper intensity gradient due to both their size and the fact that they are found within the intergranular lanes. Due to their location within the intergranular lanes, MBPs will have minimum intensity values at their boundaries. The turning points of the intensity profile are used by the code to determine the boundary of the MBP at each cross-cut across the MBP. The code then looks for similar features by looking for overlapping features in the frames before and after the detection, to track the feature over time, with key information on the MBPs (e.g. location and area) as well as a binary map of all detected features in each frame stored for future use.

The Ca wide band data was employed due to their superior cadence ($\sim$1~s compared to $\sim$33~s of the Fe~{\sc{i}} images). We employed the tracking algorithm to both the Ca wide band data and the Fe~{\sc{i}} images independently. By using both data sets to track the MBPs, we were able to identify MBPs across both data sets and ensure that the same feature does not disappear between Fe~{\sc{i}} scans. This was possible due to the superior cadence of the Ca wide band, as it allows us to ensure that the features in the Fe~{\sc{i}} scans exist for the duration that we observe them. For the Fe~{\sc{i}} scans, we employ the wing positions in Stokes I for tracking the MBPs, as this position will be a continuum position, so the MBPs will be more visible and more readily tracked by our tracking code (due to the intensity profile of the MBPs in the continuum). As the code employs intensity thresholding, we need to use the Stokes I images to track the MBPs in our Fe~{\sc{i}} images.

After these steps, when cospatial and cotemporal features were detected in both the Ca wide band and the Fe~{\sc{i}} scans, the tracked MBPs were then cross referenced to total circular polarisation maps to ensure that the tracked features were magnetic in nature and not a false detection (e.g. an exploding granule or a density enhancement at the edge of a granule). We employed the 3.5$\sigma$ level in total circular polarisation to determine if the feature was magnetic and, therefore, an MBP. Cross-referencing the detected features to the circular polarisation was a necessary step given that the intensity gradient of Fe~{\sc{i}} and the Ca wide band, are not as steep as they would be in other continuum channels (e.g. G-band) and, therefore, more non-magnetic features would begin to appear in the tracking files. Essentially, the dissociation of the CH molecule within MBPs in G-band images leads to a steeper intensity gradient across the MBP \citep{Shelyag2004}. Therefore, in the G-band, it is somewhat easier to distinguish MBPs from other small, bright features in intensity images. With our Ca wide-band data and our Fe~{\sc{i}} scans we had to relax the intensity gradient requirements of the tracking code (e.g. to those used in \citet{Keys2011} and \citet{Keys2014}) to pick up fainter MBPs. This meant that small, non-magnetic bright features (which we would classify as `false detections') were detected as well as MBPs. These false detections were then removed from our tracked feature list when we employed the total circular polarisation condition to our tracked data sets, thus, leaving only MBPs within our sample. Finally, after employing these selection criteria to the data, we excluded features in our tracking results that existed for less than 3 frames in Fe~{\sc{i}} (i.e. 99~s), so that the evolution of their properties could be analysed over time. The selection criteria resulted in 300 MBPs for our sample. 

Tracking MBPs in the MURaM simulations followed a similar approach to that of our observations. The only difference between our selection criteria with the simulations and the observations is in the utilisation of a high temporal resolution continuum passband in the verification of tracked MBPs. That is, we employed 1~s cadence Ca wide-band data with our observations to verify MBPs, however, an equivalent does not exist for our simulations. However, this step is not as beneficial for the simulations as it is with the observations as the simulations are seeing free and, therefore, less likely to have MBPs dropped between frames. The synthetic Stokes I intensity images of the MURaM simulations were tracked using the same tracking algorithm on images from the continuum position of the synthetic scan. The circular polarisation of the simulations is estimated using the synthetic Stokes scans and is cross-referenced to the tracked MBPs in the simulated intensity images to ensure that the features are magnetic. We find that this step is less necessary with the simulated data, however, we included the step to maintain parity with the methodology used with the observations. Again, we only considered MBPs that existed for at least 3 frames so that their evolution could be studied. This resulted in a sample size of 449 MBPs within the simulations. 

The Fe~{\sc{i}} scans were obtained in full Stokes spectropolarimetry mode and, therefore, inversion algorithms allowed us to glean more information on the physical parameters of the MBPs over time including magnetic field. The primary inversion code that we used in this study was the NICOLE code, which was also used to synthesise synthetic Stokes profiles for our MURaM simulations (see Section~\ref{ObsSims}). NICOLE uses a Chi-square fitting procedure incorporating response functions \citep{CoboIniesta1992} to iteratively improve on model atmospheres until post-radiative transfer profiles match observations. It includes the Zeeman effect and implements a  preconditioning approach \citep{SocasNavarro1997} for multi-level atoms. We inverted all tracked MBP pixels with three cycles with increasing nodes in temperature, LOS velocity and LOS magnetic field for each cycle. In the first inversion cycle, we had relatively few nodes, which allowed a good first approximation to be made. In subsequent cycles, the number of nodes was increased to improve the fits and to allow the code more freedom to fit more complex properties of real data, such as varying line asymmetries, which require more stratified atmospheres. This is a similar approach to the one proposed by \cite{CoboIniesta1992} and further described in  \citet{SocasNavarro2011}. The FAL-C \citep{FALC1990} quiet Sun model was used for our initial model. Low signal-to-noise in Stokes Q and U led us to exclude Stokes Q and U in our inversions. Given that the signal of Stokes~V is weaker than Stokes~I, we specified more weight to Stokes~V to ensure a better fit was achieved in the inversions. These weights were also modified between cycles to improve fits to the data. Furthermore, after rigorous tests of assorted values, as well as setting it as a free parameter, the filling factor was set to 1 for both our SIR and NICOLE inversions and was not allowed to vary between cycles. These tests were spurred by the findings of \citet{Criscuoli2014} and were applied independently for both inversion codes. We found that setting the value to 1 returned better fits and more accurate results for our dataset.

\begin{figure*}[h!]
\makebox[\linewidth]{
   \includegraphics[width=0.93\linewidth]{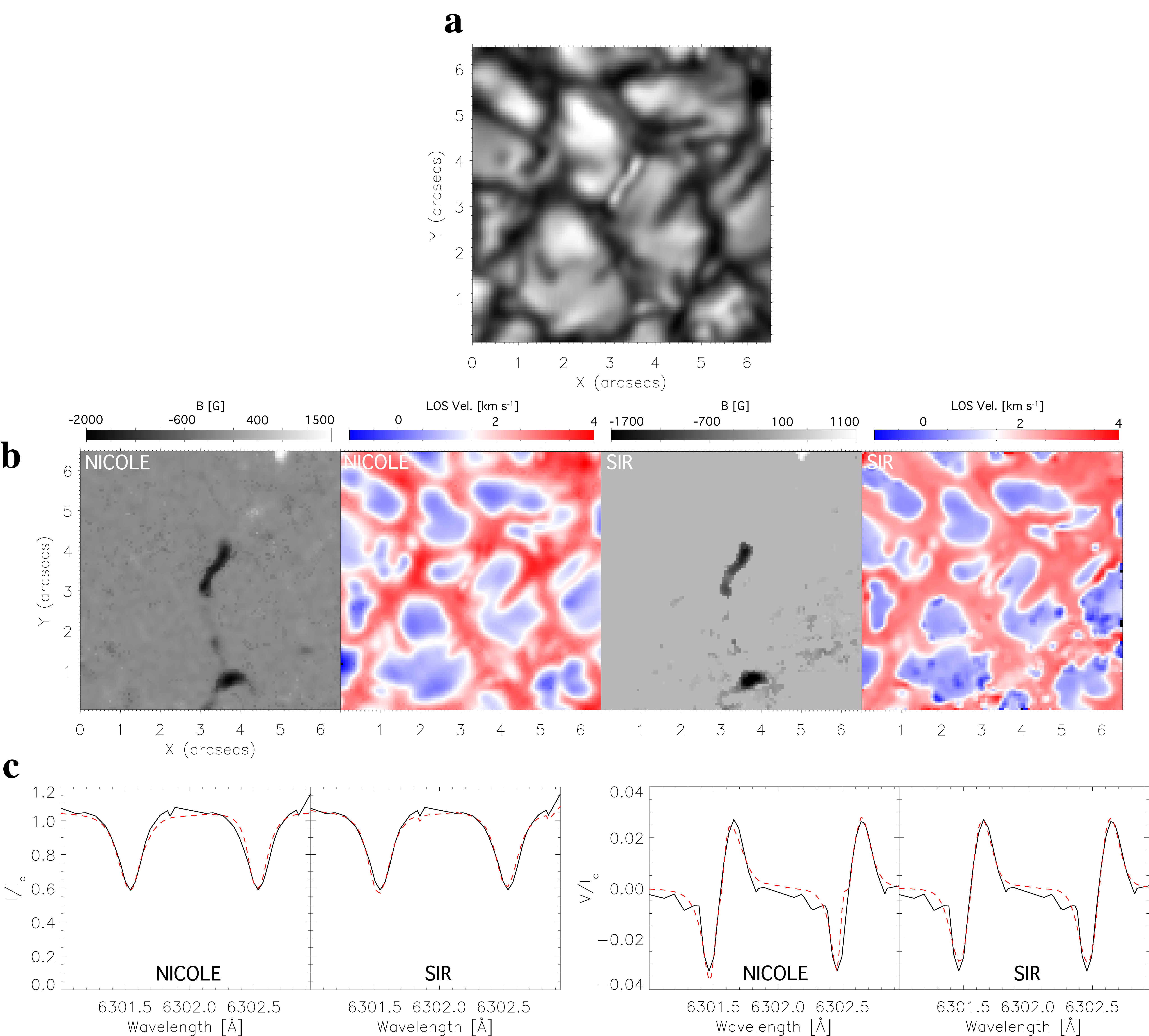}
	}
     \caption{Sample inversion outputs from NICOLE and SIR. Panel~\textbf{a} shows an MBP of interest that was inverted with both codes. The MBP is roughly central within the field-of-view, with an MBP of lower intensity also visible near the bottom of the image. Panel~\textbf{b} shows the LOS magnetic field and LOS velocity obtained with NICOLE and SIR, respectively. These are displayed for the $\log \tau=-1$ level of the atmosphere. The inversion outputs are consistent across both methods. Panel~\textbf{c} shows the fits to Stokes I (left) and Stokes~V (right). The fits are shown for a single pixel within the center of the MBP seen in the center of the FOV in panel~\textbf{a}. The black lines are the observed Stokes profile for this pixel, while the red dashed lines are the synthetic profiles after inversion. The obtained fits for both methods are fairly accurate, and given the results are consistent across both methods, we have confidence in the output models from the inversion codes.
}
     \label{Fig2}
\end{figure*}
Table~\ref{Table1} summarises the parameters used in our inversion runs and the weights applied to Stokes I, Q, U and V (relative to Stokes I). We note that in inverting the data it was necessary to interpolate the scans in Fe~{\sc{i}} onto a constant wavelength grid. Wavelength positions in the interpolated grid that do not correspond to those in the observed scans were given zero weighting. This is a necessary step for most inversion codes. Furthermore, the telluric line in the wing of the 6302~{\AA} line was masked out and was, therefore, not considered in the inversions. We also note here that both the 6301~{\AA} and 6302~{\AA} lines were inverted simultaneously under the assumption of LTE, for which NICOLE uses the approach from MULTI \citep{Carlsson1986}. The output model for each cycle was used as the input model for the subsequent cycle. A regularisation term was applied in the first cycle, i.e, a penalty is applied to the $\chi^2$ value when the model has undesirable behaviour such as large changes very localised in height, leading to a preference for smoother models when looking for the first valley in the parameter space that minimises $\chi^2$. This term was progressively reduced on subsequent cycles until it was negligible in the final cycle. 

\begin{table*}[h!]
\centering                                    
\caption{Nodes employed and weights used for Stokes parameters across inversion cycles.}         
\label{Table1}      
\begin{tabular}{l | c c c | c c c}         
\hline\hline
 & \multicolumn{3}{c|}{NICOLE}  & \multicolumn{3}{c}{SIR} \\                       
\textbf{Free Parameter} & \textbf{Cycle 1} & \textbf{Cycle 2} & \textbf{Cycle 3} & \textbf{Cycle 1} & \textbf{Cycle 2} & \textbf{Cycle 3}\\
\hline                                   
Temperature & 2 & 4 & 7 & 2 & 5 & 5 \\
LOS Velocity & 1 & 2 & 4 & 1 & 2 & 2 \\
LOS B Field & 1 & 2 & 3 & 1 & 2 & 2 \\
Inclination & 0 & 0 & 0 & 0 & 0 & 1 \\
\textbf{Weights} & & & & & & \\
\hline
Stokes I & 1 & 1 & 1 & 1 & 1 & 1 \\
Stokes Q & 0 & 0 & 0 & 0 & 0 & 0 \\
Stokes U & 0 & 0 & 0 & 0 & 0 & 0 \\
Stokes~V & 2 & 2 & 5 & 5 & 10 & 10 \\
\hline                                            
\end{tabular}
\end{table*}

We have also employed the Stokes Inversion based on Response functions \citep[SIR;][]{CoboIniesta1992,BellotRubio2003} code for a selection of MBPs (which are described in more detail in Section~\ref{Results}) to complement the NICOLE inversions. As with NICOLE, several cycles were required for SIR to match synthetic profile fits to the observed Stokes profiles, thus, minimising the differences between the two with a Marquart non-linear least-squares algorithm \citep{Press1986}. Again, we used an approach whereby the free parameters (i.e. the nodes) increase between cycles to improve the fits. Slightly fewer nodes were required to achieve a good fit with SIR in comparison to NICOLE. 

SIR uses an initial guess atmosphere based on a FAL-C model with a uniform field (750~G) and LOS velocity (1~km\,s$^{-1}$). No such magnetic field or LOS velocity was assumed for our input model with NICOLE, whereby the code will utilise an initialisation scheme where the initial guess is randomised. It is therefore likely that the minimisation needed to improve the fits were achieved faster and with fewer nodes in SIR in comparison to NICOLE. A detailed comparison of the two methods is outside the scope of this work. However, we stress that both codes were capable of achieving excellent fits to our data (see Figure~\ref{Fig2}). NICOLE was chosen as the primary inversion code for this work due to the built-in MPI parallelisation.

The parameters used for the SIR inversions are included in Table~\ref{Table1}. We note that rigorous tests on single pixels with varying degrees of polarisation in Stokes~V and from different regions (i.e. intergranular lane, MBP, and granular pixels) where employed to determine the best parameters fitting the data. These tests were carried out independently for both NICOLE and SIR. Sample fits for both codes can be seen in Figure~\ref{Fig2} as tested on the same MBP pixel. A similar plot comparing the fits in Stokes I and V for both a weak and a strong MBP can be seen in \citet{Keys2019} Figure~2.

				
\section{Results \& discussion}
\label{Results}

In this work, we examine the temporal evolution of various MBPs parameters, with emphasis on what we define as the `strong' group of MBPs. We define a strong MBP as one which has an excursion above 1100~G at least once during its lifetime. The choice of the 1100~G threshold is evident when considering the work of \citet{Keys2019}, in that this was roughly the lower limit separation between the groups in the bimodal distribution of B-field strengths. This selection criteria reduced the sample to 64 MBPs. An example of one such MBP can be seen in Figure~\ref{Fig2} with panel \textbf{b} showing the LOS B-field and LOS velocity as obtained with NICOLE and SIR for that particular frame. Note, that the values for inverted parameters discussed within this section were taken as an average from the inversions over optical depths (log$\, \tau$) ranging from $-1.5$ to $-0.5$.

When we examine closely the temporal evolution of the B-field we find that MBPs do not belong to the strong group for their entire lifetime. That is, an MBP can traverse between the weak and strong groups multiple times. The difference between the properties that we report here and those of previous studies is that the magnetic field amplification is rapid (roughly doubles in $\sim$30~--~100~s) often relaxing back to typical weak group B-fields as well. Also, these fast amplifications/relaxations can occur at multiple instances in the MBP lifetime, not unlike behaviour we would expect for an MBP supporting an MHD wave mode \citep{FujimuraTsuneta2009}. 

We also employed 200~G MURaM simulations to investigate if the amplification processes can be reproduced in the simulated MBPs. An  analysis similar to the observations was carried out with the simulations with the MBPs tracked throughout the duration of the time series ($\sim$90~minutes of data with a cadence of around 17~s) before analysing properties such as LOS B-field and velocity. These properties  do not rely on the inversion of Stokes parameters. In total we found 449 MBPs in the simulated dataset. We note that we do not find the same bimodal distribution as we do for our observations in these simulations \citep{Keys2019}.

As there was no discernible bimodal distribution in the simulations, we employed the 1100~G LOS B-field value found in the observations as the separator between weak and strong MBPs. We chose this threshold to select those MBPs that crossed between groups at least once in their lifetime, while existing for several frames below this value as well. Again, we find B-field amplifications that occur on short timescales due to the same reasons as in our observations. With the simulations we find an additional amplification process that did not appear in our observations, namely, amplification due to vorticity. 

Upon closer inspection, we can attribute the rapid amplification in our observations to three different scenarios, namely, (1) convective collapse, (2) MBP compression due to granular expansion, and (3) MBP merger events. Within the following subsections we will describe each of these processes in turn for both our observations and simulations.

\subsection{Convective collapse}
\label{Collapse}

\begin{figure*}[h!]
\makebox[\linewidth]{
   \includegraphics[width=0.9\linewidth]{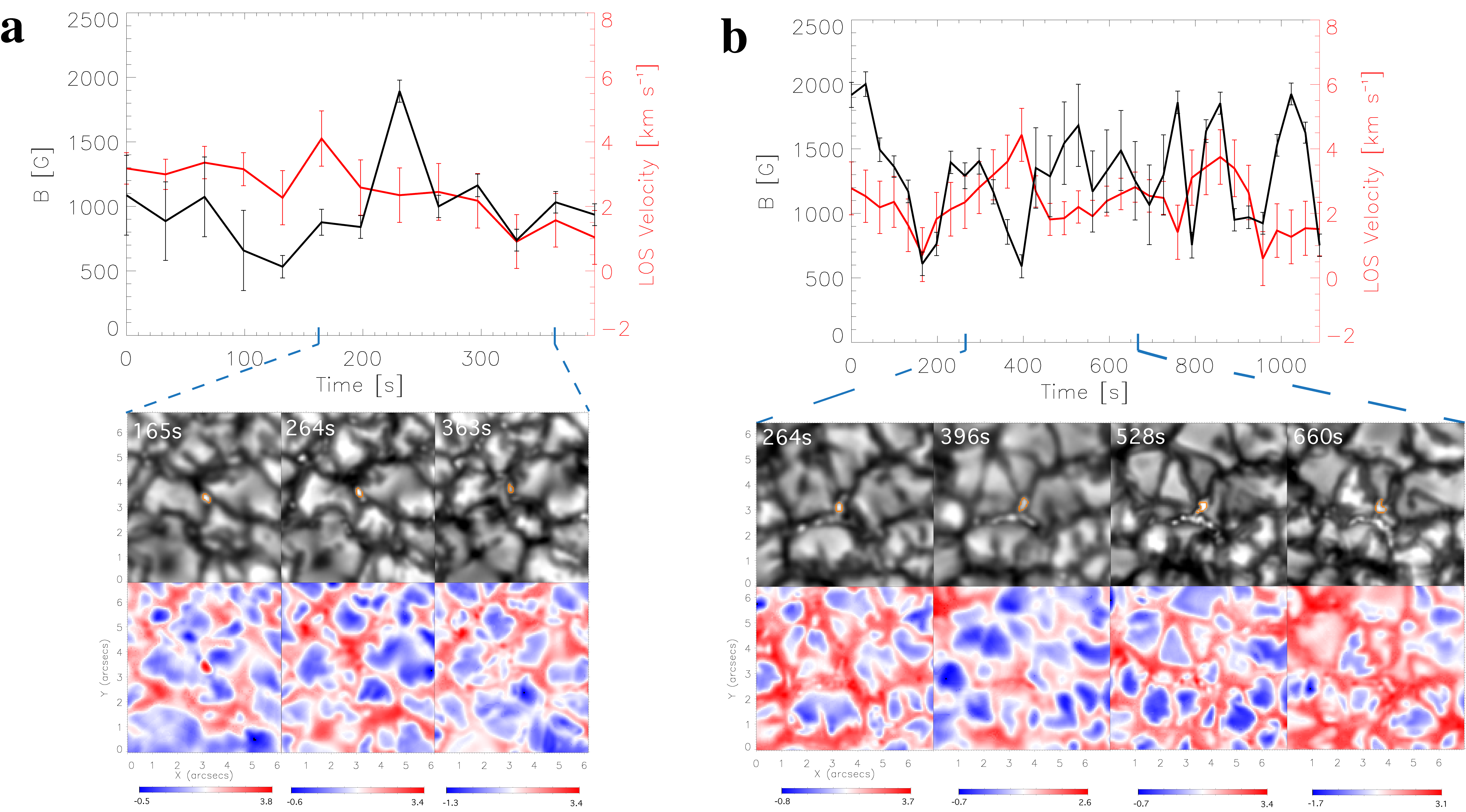}
	}
     \caption{Examples of the temporal evolution of MBPs undergoing convective collapse. Within the plots, the black line indicates the LOS magnetic field strength, while the red line indicates the LOS velocity in each frame that the MBP was detected. The values were obtained from NICOLE inversions. The calculation of error bars follows \citet{Reid2016}. In each panel, below the temporal evolution plot, we show intensity images (taken at 6301.0392~{\AA} in our CRISP line scans) and LOS velocity maps (from NICOLE inversions) for the MBP at key stages in its evolution. Blue dashed lines indicate the time interval which these images are taken. Orange contours in the intensity images show the MBP, as tracked by our algorithm, used to make the evolution plots above. Panel~{\textbf{a}} shows a simple convective collapse case study. Here the LOS velocity within the MBP increases just prior to the amplification of the magnetic. The intensity images show the MBP getting smaller and brighter during this process. Panel~{\textbf{b}} shows the case for multiple collapse events within an MBPs lifetime. This plot shows the complexity of MBP evolution, with collapse events occurring at multiple times in the MBP's lifetime. More details on both these MBPs can be found in the main text.
}
     \label{Fig3}
\end{figure*}

The process of convective collapse is characterised by a peak in LOS velocity observed a few frames prior to the B-field amplification (e.g. see Figure~\ref{Fig3}). Figure~\ref{Fig3}a shows an example of convective collapse, similar to those seen previously in the literature. We note that these values, and those within all subsequent figures, are taken as an average over a small 3$\times$3 pixel box about the barycentre of the MBP. Similar plots are seen when considering an average over the MBP as a whole, however, the clarity of the peaks is reduced when weaker edge pixels are considered. We chose a box, so as to smooth out any inconsistencies in the inversions for single pixels. We kept the box relatively small to account for changes in shape of the MBPs over time (e.g. when the MBP becomes more elliptical). 

For the convective collapse example seen in Figure~\ref{Fig3}a, the downflow within the MBP increases from 2.23$\pm$0.89~km\,s{$^{-1}$} to 4.16$\pm$0.90~km\,s{$^{-1}$} over 33~s, before relaxing again to 2.56$\pm$0.90~km\,s{$^{-1}$}. This peak in the LOS velocity precedes the peak in the B-field by  66~s. The B-field increases from 858$\pm$53~G at the time of the LOS velocity peak, to 1926$\pm$45~G at the time of the maximum B-field value, before relaxing back down to 987$\pm$45~G 33~s later. The B-field effectively doubles as it peaks. The area of the MBP reduces by 23\% from the time of peak LOS velocity to the time of the B-field peak. We also note that the intensity of the MBP, (not shown in the plot for the clarity of other parameters, but visible in the intensity images included below the plot), peaks with the B-field peak as expected. 

\begin{figure}[h!]
\makebox[\linewidth]{
   \includegraphics[width=0.9\linewidth]{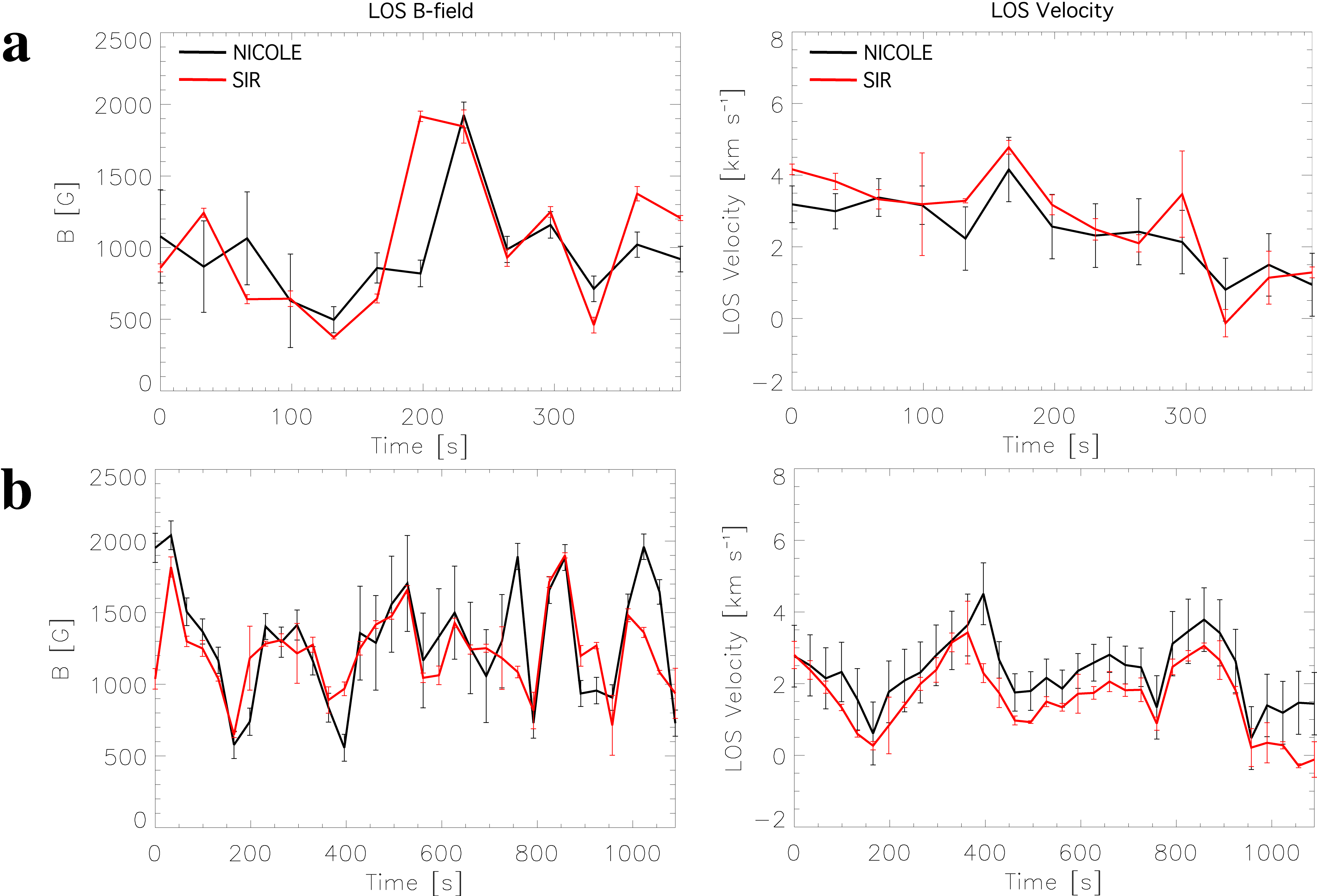}
	}
     \caption{Comparisons of the results obtained with NICOLE and SIR for the LOS magnetic field (left column) and LOS velocities (right column) for the MBPs displayed in Figure~\ref{Fig3}. Row {\textbf{a}} here corresponds to the MBP in panel {\textbf{a}} in Figure~\ref{Fig3}, while row {\textbf{b}} corresponds to the MBP in panel {\textbf{b}}. In all plots the black line indicates the results from NICOLE while the red line shows the results from SIR. The error bars in NICOLE values follow the approach of \citet{Reid2016}, while the error bars with SIR parameters were calculated with the approach described in \citet{SocasNavarro2011} Section~5. On the whole, both methods return very similar evolutionary characteristics, often displaying the same peaks and troughs in the parameters. For the most part, the values obtained for both fall within their respective error bars. Discrepancies between the two approaches could be due to the success of fitting in a given frame for a particular technique. The results with these two independent techniques, however, show consistency in our results for the evolutionary characteristics of the MBPs.
}
     \label{Fig4}
\end{figure}

We find similar values and timescales for this MBP when using SIR (see Figure~\ref{Fig4} for a comparison between NICOLE and SIR inversions). The LOS velocity peaks at the same time with SIR and increases from 3.28$\pm$0.06~km\,s{$^{-1}$} to 4.78$\pm$0.20~km\,s{$^{-1}$} before relaxing to 3.17$\pm$0.28~km\,s{$^{-1}$} over the same timescale, then decreases further to 2.10$\pm$0.24~km\,s{$^{-1}$} after the B-field relaxes. The B-field rises from 645$\pm$31~G at the time of the LOS velocity peak, to a peak of 1916$\pm$36~G 33~s later. The B-field drops to 930$\pm$61~G 66~s later. Compared to the results from NICOLE, the increase in the B-field found with SIR is slightly larger (magnified by $\sim$3 times as opposed to $\sim$2.2 times) and occurs in a shorter timescale (33~s as opposed to 66~s). With SIR, the magnetic field remains high for an extra frame although they both relax back to lower B-field values over the same time. 

In most instances, the two codes return very similar values for magnetic field (at least within their respective errors). We looked closer at this case, where the magnetic field with SIR does not quite match that of NICOLE. The issue appears to be due to the success in fitting the Stokes~V profiles in this particular frame. With SIR, the synthetic Stokes~V fits are not quite as accurate as the corresponding NICOLE fits. In this instance, the lobes appear to be split further than they actually are in the observations in the synthetic Stokes~V profiles output from SIR. As such, the estimation of the LOS B-field will be larger. The fits with NICOLE are closer to the observed profiles, and back of the envelope estimations of the B-field from Zeeman splitting suggests that, in this case, NICOLE returns the more accurate value. In general, the few cases were this sort of discrepancy occurs in our data is mostly due to poor fits in that particular frame. The fits are more accurate in other frames with SIR than in this particular frame, which suggests that this is something of an anomaly. Also, we note that in most cases where these discrepancies exist, NICOLE was more successful in accurately fitting the Stokes~V profiles. We do not have a definitive reason for this, though it is probable that the regularisation term and a greater number of nodes in LOS B-field (Table~\ref{Table1}) with NICOLE gives it more freedom to fit parameters, and achieve a more accurate fit of the profiles.

We also note that there is a slightly lower increase in LOS velocity values over the peak in LOS velocity in the case of SIR. It is unlikely that both codes would ever return the exact same values, however, the change in values for both follows the same trends with similar amplification values. Also, the uncertainties associated with the measurements for both the NICOLE and SIR methods indicate that these rapid increases are real as opposed to errors in the inversions. The uncertainties for both methods suggest that the values found for the peak values are consistent across methods, which is reassuring in our selection of free parameters and weighting in our inversions (see Section~\ref{Methods}). On the whole, the results for each of these MBPs is consistent between NICOLE and SIR, as can be seen in Figure~\ref{Fig4}. 

We note that with SIR, the uncertainties are calculated from the response functions for the inversions using the approach described in \citet{SocasNavarro2011} Section~5. With NICOLE, it is decidedly more difficult to calculate the response functions as one needs to make changes within dependencies in the code. We decided to avoid doing this so that we could use the NICOLE distribution as it was intended by the developer. To work out the errors in the physical values calculated with NICOLE, we used the technique employed by \citet{Reid2016}, whereby, the inversions were run on a 150$\times$150 pixel$^2$ patch of quiet Sun devoid of MBPs (or any magnetic features) using the weighting and free parameters used in our inversions of the MBPs pixels for each frame of our observations. The variance of all pixels in this patch was then used to calculate the uncertainties for a given model parameter at a given optical depth. In the case of the plots shown for individual MBPs inverted with NICOLE, this was taken as a mean value between $\log \tau = -1.5$ to $\log \tau = -0.5$. This was done for each frame in order to evaluate how variations due to seeing etc. affected the calculation of these values. The uncertainties given in this section for NICOLE and SIR outputs are displayed based on the values found using this approach. 

The increase in LOS velocity does not necessarily happen prior to the formation of the MBP, but can occur at any stage during its life cycle. We find that the process can occur at multiple occasions during the MBPs lifetime. This can be seen in Figure~\ref{Fig4}b. The LOS velocity peaks at 3.31$\pm$0.87~km\,s{$^{-1}$} 66~s before the initial B-field peak when the MBP is first detected. The B-field increases from 381$\pm$51~G before the MBP is initially detected by our tracking code to a maximum of 2040$\pm$50~G just after it is initially detected. Convective collapse also appears to occur again in the time frame around 260~s to 530~s, after the MBP is first detected. The LOS velocity increases from 2.31$\pm$0.86~km\,s{$^{-1}$} 266~s after first detection, peaking at 4.51$\pm$0.53~km\,s{$^{-1}$} at around 397~s after first detection. The B-field rises from 557$\pm$47~G at the peak in LOS velocity and rises to 1704$\pm$160~G 132~s after the initial peak in LOS velocity. The average time observed between LOS velocity peak and B-field peak was around 100~s for the MBPs in our sample that displayed signatures of convective collapse. This is consistent with previous work on convective collapse which leads us to believe that the physical process is the same between these traditional collapse events and those we describe here. These double collapse events have been observed before see, for example, \citet{Utz2014} Section~4.1 for an example from the literature.

\begin{figure}[h!]
\makebox[\linewidth]{
   \includegraphics[width=0.9\linewidth]{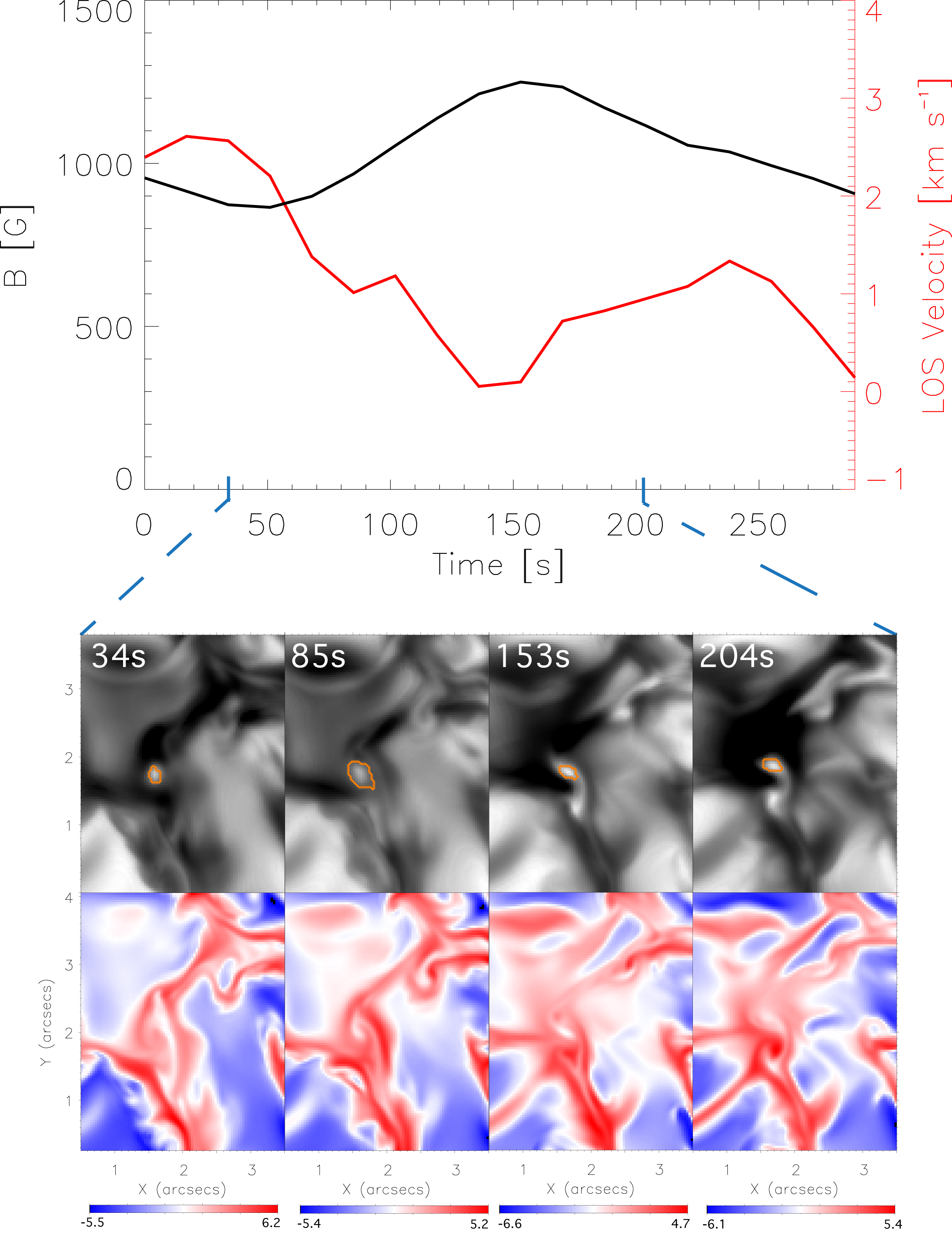}
	}
     \caption{Example of convective collapse from the MURaM simulations. The top plot shows the evolution of the magnetic field in the vertical direction (black) with the evolution of the LOS velocity (red), where positive velocity values indicate a downflow. The panels below show the evolution of the MBP in intensity (top rows) and LOS velocity (bottom row) for the time frame indicated by the blue dashed lines. Orange contours in the intensity images show the MBP, as tracked by our algorithm, used to make the evolution plots above. The LOS velocity peaks $\sim$136~s prior to the peak in magnetic field. It can be seen in the intensity images that the MBP reduces in size in this period of time and the intensity increases, which is indicative of convective collapse.
}
     \label{Fig5}
\end{figure}

Figure~\ref{Fig5} shows an example of convective collapse in the simulations. The B-field rises from 870~G to 1274~G in 102~s, and remains above the 1100~G threshold for 68~s. The LOS velocity peaks with a downflow of 2.67~km\,s$^{-1}$ just after the MBP is first detected, and the time between the peak in LOS velocity and the subsequent peak in B-field is 136~s. The LOS velocity at the time of peak magnetic field is a downflow of 0.02~km\,s$^{-1}$. The B-field reduced to below the 1100~G threshold over a period of 136~s before the MBP disappears. These timescales are close to others found in the literature \citep{Nagata2008,Hewitt2014}. Also note that the area of the MBP decreases by about 33\% in this process, which is close to values reported by \citet{Fischer2009} ($\sim$20\% reduction).

\subsection{Granular compression}
\label{compression}

\begin{figure*}[h!]
\makebox[\linewidth]{
   \includegraphics[width=0.9\linewidth]{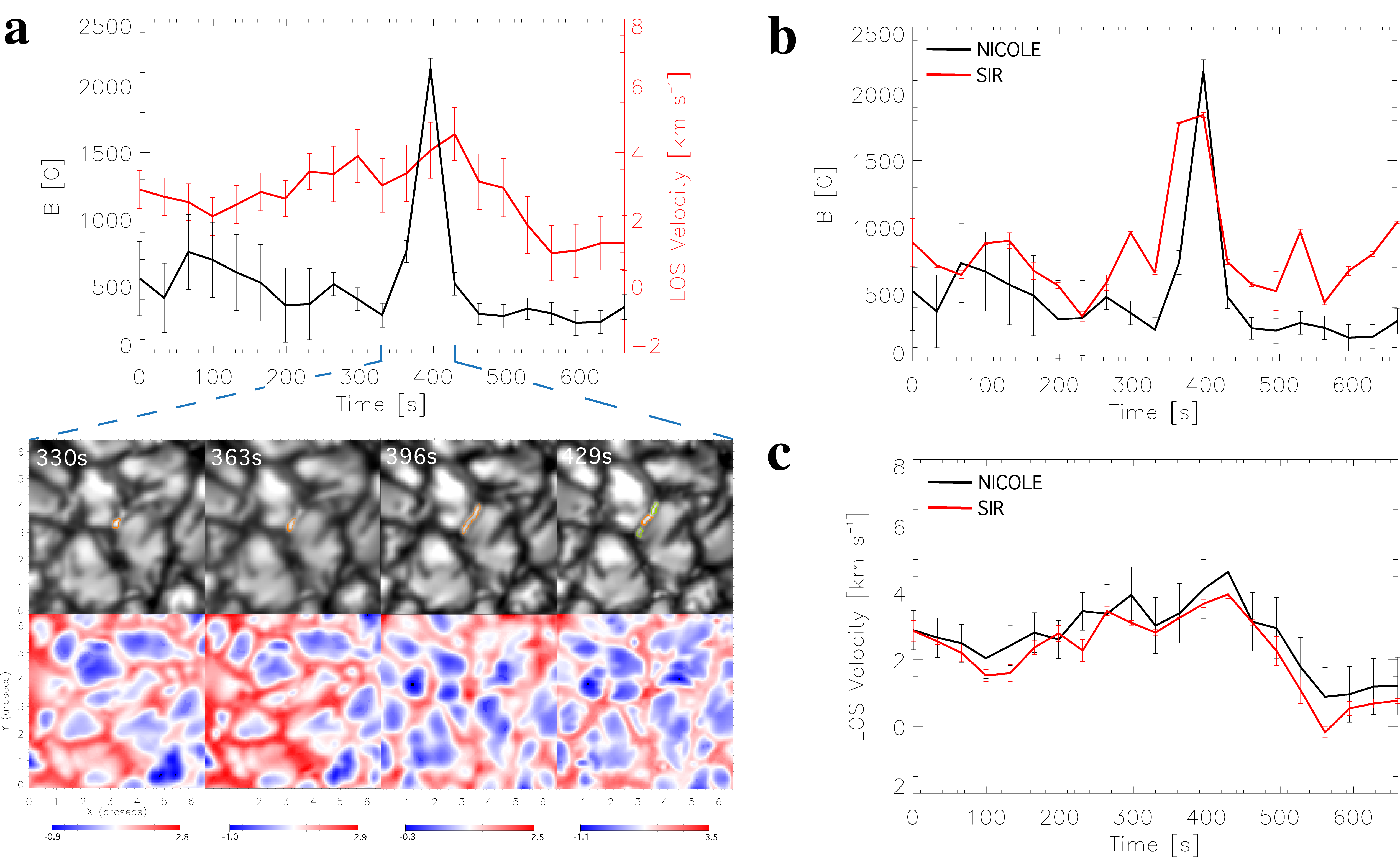}
	}
     \caption{Example from our observations of granular compression leading to magnetic field amplification in an MBP. Panel {\textbf{a}} shows the temporal evolution of the MBP, with the plots at the top showing the output of the NICOLE inversions, with the black line indicating the LOS magnetic field and the red line the evolution of the LOS velocity. The images below this plot show the evolution of the MBP during the amplification phase indicated by the blue dashed lines. Orange contours in the intensity images show the MBP, as tracked by our algorithm, used to make the evolution plots above. Green contours in the final panel show the MBPs that split from the main MBP under investigation after the MBP is compressed and stretched out. LCT analysis of the granular flow direction show that the granules expand towards the MBP, resulting in its compression and the amplification of the magnetic field. The LOS velocity peaks after the magnetic field peak, contrary to the standard convective collapse process. Panel {\textbf{b}} and {\textbf{c}} show comparisons between the temporal evolution of the LOS magnetic field and LOS velocity, respectively, as derived from the NICOLE (black) and SIR (red) inversions. The two methods show consistent results.
}
     \label{Fig6}
\end{figure*}

As the granules evolve, MBPs get jostled and change shape as they are subjected to external plasma pressure. We find several instances where the MBP is effectively `squeezed' leading to an amplification of the B-field. In Figure~\ref{Fig6}a, we show a case where this occurs between times 363~s and 429~s. The movements of the granules on either side of the MBP cause it to get compressed. We used Local Correlation Tracking \citep[LCT;][]{NovemberSimon1988} of the granular evolution in tandem with difference imaging of the granules, to quantify that the direction of granular flow changes just prior to the B-field amplification. It is the direction of the flow that acts to compress the MBP. Note that with convective collapse we expect to see a compression (the collapse) of the MBP with magnetic field amplification, however, here the downflow peaks after the magnetic field amplification. The LOS velocity steadily rises from $\sim$2.04~km\,s{$^{-1}$} 330~s before the peak in B-field, to a maximum of 4.63~km\,s{$^{-1}$} 33~s after the magnetic field has peaked. The B-field goes from a minimum of 235~G before rapidly peaking at a B-field of 2172~G 66~s later before dropping to 480~G 33~s after this peak. 

The behaviour of the LOS velocity suggests that the phenomena leading to the field amplification may be different. This is supported by our LCT analysis. In this case, as the granules evolve, the compression is relaxed and the B-field drops down below a kilogauss again. This process is clearly distinct from convective collapse. Prior to the amplification in the B-field within the MBP, the granules on either side move towards the same direction (towards the upper left of the subfield shown in the intensity images of Figure~\ref{Fig6}a) at a velocity of 0.9~km\,s{$^{-1}$} (to the right of the MBP) and 0.3~km\,s{$^{-1}$} (top left beside the MBP). Just as the MBP amplification occurs, the granule above and to the left of the MBP stops moving while the granule to the right side of the MBP accelerates slightly to 1.2~km\,s{$^{-1}$}. After the amplification, the granule to the top left of the MBP remains stationary, while the granule to the right side of the MBP drops to 0.6~km\,s{$^{-1}$} and changes direction slightly, and is no longer pointed towards the MBP. The distance across the intergranular lane decreases from $\sim$450~km to $\sim$300~km as the magnetic field amplifies. The MBP gets stretched out at this time and eventually the shift in direction of the granular flow to the right of the MBP causes it to split, which is  coincident with a reduction in the B-field.

Although there are  some similarities with convective collapse, the processes that lead to the field amplification are sufficiently distinct. In this case the field appears to increase as the granules expand, squeezing the MBP. Now it is possible, and it can be seen in our plot of the LOS velocity, that the LOS velocity increases with the granular expansion and the decrease in width of the intergranular lane, which could result in the increase in field strength within the MBP here. However, the increase in LOS velocity does not occur without the expansion of the granules and, unlike the standard convective collapse model, the LOS velocity does not peak prior to the B-field peak here, which would suggest that the peak here is due to the granular expansion. The disintegration of the MBP in this case is then instigated by the granular evolution as a sort of horizontal shear flow appears to split the MBP. 

Similar behaviour of granular motion leading to B-field amplification is observed in other MBPs too, which leads us to believe that this process is distinct from convective collapse. The example shown in Figure~\ref{Fig6} is given due to its relative simplicity. In another case, we observe the MBP to move rapidly between two granules before it meets the wall of a third granule. This causes the MBP to stop abruptly and compress, which is coincident with a rapid rise in the B-field. The MBP then relaxes as the external forces settle and the B-field drops. In another case granular expansion again leads to a rise in the B-field, however, on this occasion the MBP  is effectively `pinched' as it stretches out causing it to split in two. Upon splitting, the B-field of the two features that appear from the original MBP drops to below kilogauss values. In many cases where there is rapid amplification due to granular motion, we see an increase in gas pressure within the MBP at or very close to the peak in B-field (though not always). Of the 64 strong MBPs in our sample, at least 18 show unambiguous evidence for rapid B-field amplification due to granular compression at least once in their lifetime. There are several more cases where rapid amplification is due to granular compression although not as evident as the 18 MBPs mentioned above. Therefore, the process of granular forcing leading to rapid B-field amplification appears to have a significant impact on the evolutionary properties of MBPs.

It should be noted here that external forcing from granules is a key component for driving sausage modes in solar magnetic structures \citep[][]{Dorotovic2008, Grant2015, Freij2016, Keys2018}, and it is possible that upwardly propagating wave phenomena are associated with these compression events. Analysis of such phenomena in these MBPs will be considered in a future publication and is outside the scope of this work. The nature of compression would suggest that the excitation of sausage modes in MBPs is significantly different to those described in the studies of pores outlined above. Pores exist on timescales of hours whereas MBPs exist on timescales of several minutes. The longer lifetimes, and less dynamic behaviour of pores means that they experience continual forcing from expanding granules, which drive the oscillations. For MBPs, the forcing from the granules may cause the MBP to be compressed, however, more often the granule evolution causes the MBP to move in a transverse motion (possibly inciting a kink mode) or causes the MBP to split apart. Therefore, sausage mode studies in MBPs may be better suited to longer lived network bright points, which are trapped by supergranular flows and, consequently, subjected to external granular forcing more frequently. 

Furthermore, sausage mode studies frequently examine the change in cross-sectional area of the feature as a signature of the wave mode. In our examples of granular compression in the MBPs, the cross-sectional area does not necessarily change on compression. In the example shown in Figure~\ref{Fig6}, the area of the MBP changes from 19 pixels before amplification, to 22 pixels at peak B-field, to 10 pixels when it has relaxed. This is not a significant change from just prior to the amplification and at the moment of highest B-field. The area drops considerably between the point of highest B-field and the frame after as the MBP is considered to have split. In general, as MBPs are quite dynamic, compression can lead to the MBP becoming extended into a more elongated structure. This may not necessarily change the number of pixels that are detected as being part of the MBP.  

In this case, whereby the MBP is elongated by granular compression, problems may arise if a circular geometry is assumed. Instead, a more nuanced approach is required, such as considering each MBP individually, or at least modelling them as ellipsoidal in shape. These effects of varying MBP geometry are also considered by \citet{VanKooten2017} using high resolution MURaM simulations. Considering the MBP in this example as an ellipse, the semi-major axis varies from five pixels to nine pixels then finally to five pixels across the time frame of the amplification. The semi-minor axis of the MBP varies from four pixels just before the peak to three pixels at the peak before returning to three pixels just after the peak. The eccentricity of the MBP varies from 0.6 before the peak to 0.94 at the peak to 0.8 after the peak. Clearly, looking at the semi-major axis variation shows the compression of the MBP, whereas if we treat the MBP as a circle (where the corresponding radius would vary from roughly 125~km, to 130~km to 90~km across the amplification) this information is lost. Therefore, we urge caution in large-scale statistical studies of waves in MBPs due to complexities in their evolution, and perhaps a different approach is required for large scale studies of sausage modes in MBPs \citep{Jafarzadeh2017a}.

\begin{figure}[h!]
\makebox[\linewidth]{
   \includegraphics[width=0.9\linewidth]{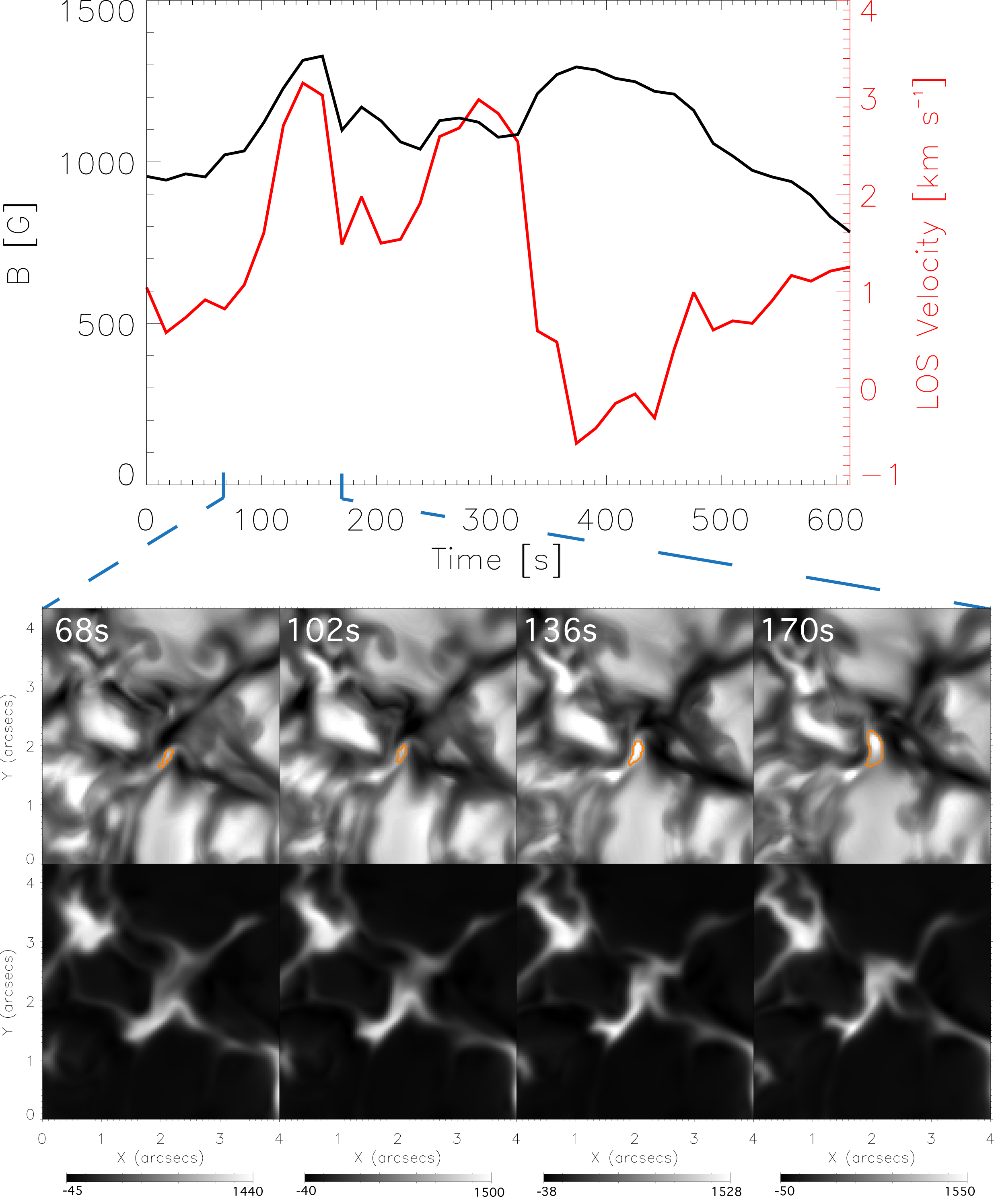}
	}
     \caption{Example from our MURaM simulations of granular compression leading to magnetic field amplification in an MBP. The black line shows the magnetic field in the vertical direction while the red line indicates the LOS velocity. The panels below show the evolution of the MBP in the intensity images (top) and the evolution of the magnetic field for the corresponding intensity images, for the time period indicated by the blue dashed line. Orange contours in the intensity images show the MBP, as tracked by our algorithm, used to make the evolution plots above. Analysis of the horizontal velocity of the surrounding granules shows that the granules expand and compress the MBP resulting in an amplification of the magnetic field (the first peak in the top plot). The LOS velocity peaks with the peak in magnetic field. The plots of the magnetic field in the region over this period of time show the compression of the magnetic flux within the intergranular lanes.
}
     \label{Fig7}
\end{figure}

Figure~\ref{Fig7} shows an example of compression leading to magnetic field amplification (in the first magnetic field peak) from our MURaM simulations. The magnetic field rises from 963~G to 1356~G over 119~s. The LOS velocity peak coincides almost exactly with the peak in B-field (going from 0.7~km\,s$^{-1}$ to 3.2~km\,s$^{-1}$), which hints  that this is not due to the usual convective collapse process. An examination of the  LOS velocity at the location the MBP prior to its detection, does not show a sharp downflow prior to the magnetic field amplification. The temporal evolution of the MBP over this time frame, however, seems to indicate that the amplification in magnetic field is due to compression. The intergranular lane decreases in size by 30\% from the start of the MBP lifetime to the time of the peak B-field. We note that, similar to  observations, the area of the MBP does not actually change in this process. As in the observations, the MBP becomes more elliptical during the amplification phase with the near circular MBP at the minimum B-field having a diameter of 200~km, while the roughly elliptically shaped MBP geometry at the peak has a semi-major axis of 375~km and a  semi-minor axis of 125~km. Effectively the MBP gets compressed from the sides, and squeezes out into a more elliptical shape governed by the now narrower dimensions of the intergranular lane. When the magnetic field of the MBP drops below the 1100~G threshold, it returns again to a roughly circular geometry with a diameter of 175~km. This is remarkably similar to the observational event, and reiterates the complex MBP evolution and that amplification due to compression needs to be considered when searching for sausage modes in MBPs.

\subsection{Merger events}
\label{Merger}

\begin{figure*}[h!]
\makebox[\linewidth]{
   \includegraphics[width=0.9\linewidth]{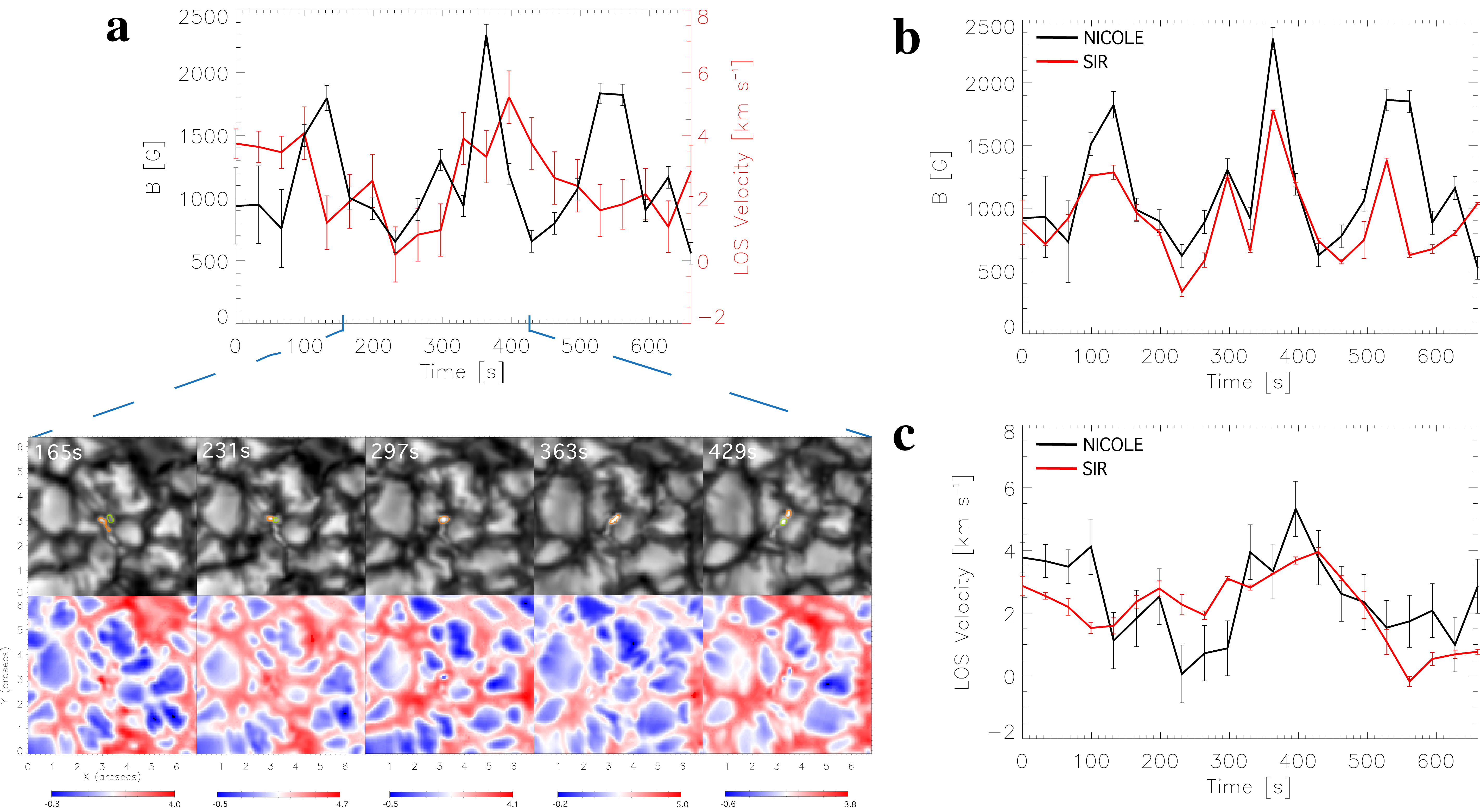}
	}
     \caption{Same plots as Figure~\ref{Fig6}, but for observations of merging MBPs leading to magnetic field amplification. The temporal evolution of the magnetic field shows how complicated MBP evolution is due to various factors. The intensity images in panel {\textbf{a}} help demonstrate this complex behaviour. Orange contours in the intensity images show the MBP, as tracked by our algorithm, used to make the evolution plots above. Green contours indicate the other MBP involved in the merger. The first small peak in magnetic field (at $\sim$1300~G) is due to a merging of two MBPs. The intensity images show the motion of several MBPs within a small group. One MBP (orange contours) is pushed by the granules past the small group, before being forced back and eventually merging with another MBP (green contour). The merger results in a peak in the LOS magnetic field. The granules continue to buffet the newly merged MBP compressing it, and resulting in the secondary peak a couple of seconds after the merger peak. The comparison between NICOLE and SIR in panels {\textbf{b}} and {\textbf{c}} confirm the amplification due to these processes.
}
     \label{Fig8}
\end{figure*}

\begin{figure}[h!]
\makebox[\linewidth]{
   \includegraphics[width=0.9\linewidth]{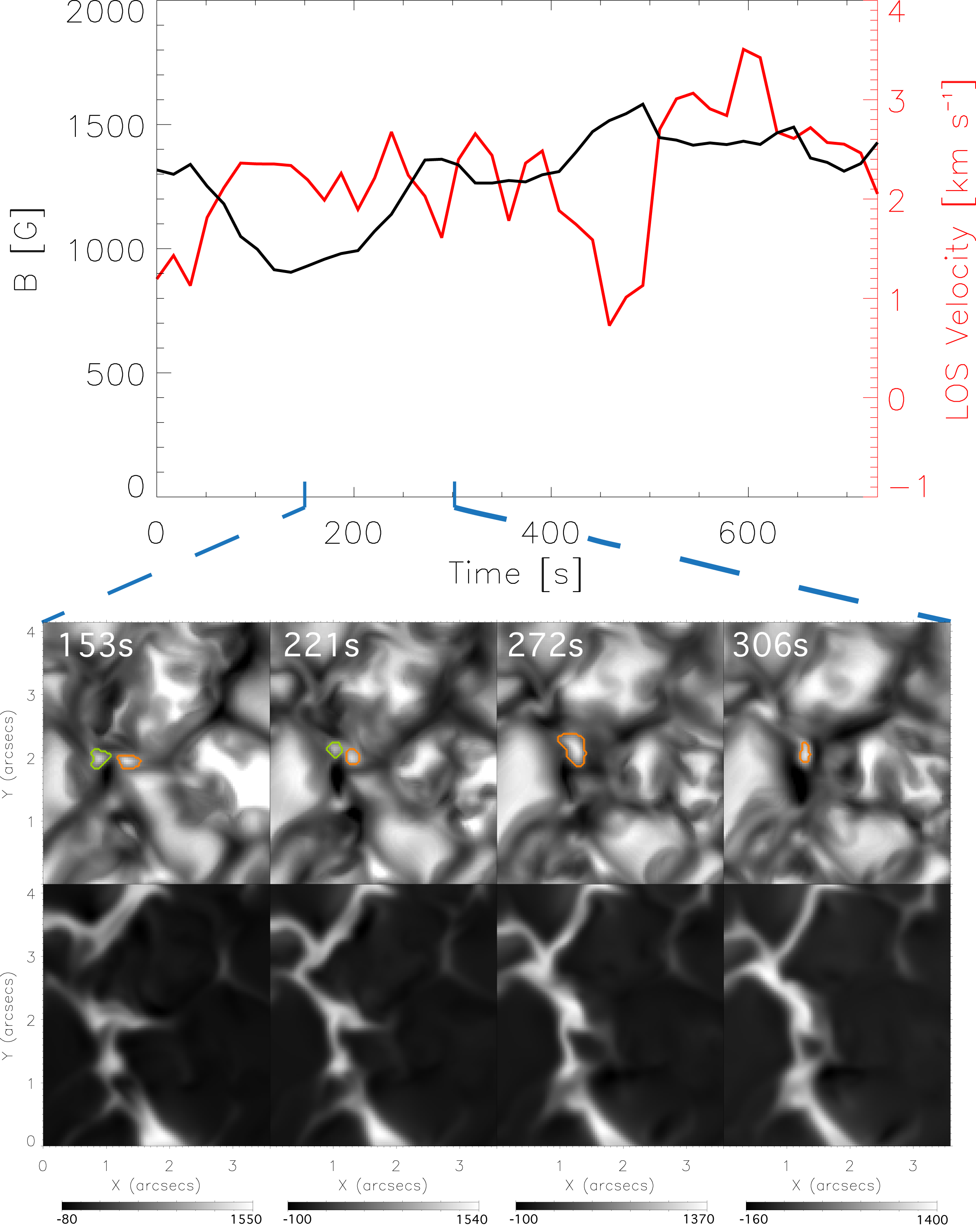}
	}
     \caption{Example of merging MBPs from the MURaM simulations that leading to magnetic field amplification. The black line shows the magnetic field in the vertical direction while the red line indicates the LOS velocity. The panels below show the evolution of the MBP in the intensity images (top) and the evolution of the magnetic field for the corresponding intensity images, for the time period indicated by the blue dashed line. Two MBPs (orange and green contours) can be seen around 1$''$ to 1.5$''$ in $x$ and 2$''$  in $y$ in the intensity images in the first frame. The two MBPs move towards each other before coalescing, with the LOS magnetic field rising and peaking as they merge. As before, the magnetic field and LOS velocity plots above represent the evolution of the orange contoured MBP.
}
     \label{Fig9}
\end{figure}

Our observations show that magnetic field amplification in MBPs can happen as a result of merging. Figure~\ref{Fig8}a shows an example for such an event where the temporal evolution of the B-field and LOS velocity shows the complexity of the MBP properties over time. It should be noted here that the plots for B-field and LOS velocity in this figure (and in Figure~\ref{Fig9} for the simulated merger event) are for the MBP that we consider to be the `dominant' MBP in the merger, that is, the MBP that the detection algorithm tracked for the longest time period prior to the merger event. Within both Figure~\ref{Fig8} and Figure~\ref{Fig9} this corresponds to the MBP outlined with the orange contour, while the other MBP involved in the merger is outlined with green contours. Defining which MBP is `dominant' in a merger is rather arbitrary and defining if the feature that forms after two MBPs merge is a new feature or a continuation of an existing feature, is a subject of debate with regards to MBPs. These definitions are somewhat outside the scope of this work as we are interested in the variation of the properties of these features when two merge to form one single entity. Regardless, the flux concentration in the lanes giving rise to the MBPs is still there along with an actual MBP in the intensity images, which our tracking code was able to detect. In Figure~\ref{Fig8}, the first peak between 66~s and 132~s is  due to convective collapse. The second peak between 231~s and 330~s coincides with the merger of two separate MBPs. Finally, the peak between 396~s and 425~s appears to be due to compression from granular motion.

The peak that we are most interested in here is the second peak, that is, the merger event. At the minimum just before the peak, the B-field is 620$\pm$91~G, while the MBP that it eventually merges with has a B-field of 475$\pm$91~G. In the next frame, 33~s later, it begins to merge with the MBP just to the right of it. At this point the B-field of the MBP has increased to 890$\pm$92~G, while the B-field of the MBP to the right of the first MBP also increases to 800$\pm$92~G. In the next frame, the two MBPs have completely merged and the B-field of the newly merged MBP is now 1306$\pm$88~G. The next frame after the merger has complete, the MBP relaxes a little and the B-field drops down to 920$\pm$88~G, before the expanding granules that resulted in the merger, compress the MBP at 363~s after it initial detection, causing the B-field to peak at 2353$\pm$88~G. In the following frame, the MBP is stretched out between the granules before it starts to split at 429~s after initial detection and the B-field of the MBP drops again. This process can be seen in Figure~\ref{Fig8}, which shows the motion of the MBPs during this whole merge and split process, and again highlights the complexity of MBP evolution.

The evolutionary track of the MBP, as the two move towards each other, appears to have some degree of rotational motion, albeit slight. The MBP in the lower part of the image at 163~s, moves up towards the other group above it, but it moves to the north west of the FOV, moving past the other group of MBPs. Around 231~s, the north westerly motion is halted by the motion of a granule, and the MBP appears to rotate in a clockwise direction due to the granule forcing it, and begins to move south east. At about 297~s after its first detection, the MBP merges with the MBP below it, when the field amplifies and the MBP appears to be more compact. The rotational motion that leads to the merger may be due to a vortex, however, we have not looked for vortical motions in the images, so we can not say for certain that this is the case. This will be the subject of future work employing techniques, such as those described in \citet{Giagkiozis2018} and \citet{Liu2019}. However, it should be pointed out here that it is the point of merging that the B-field of the MBP amplifies.

Magnetic field amplification due to merging MBPs can also be seen in the simulations and is displayed in Figure~\ref{Fig9}, specifically the second peak in the B-field plot. The B-field rises from 900~G to 1375~G in 136~s. The LOS velocity appears to fluctuate between about 1.5~km\,s$^{-1}$ and 2.5~km\,s$^{-1}$ just before and during the amplification peak in magnetic field. An examination of the MBP and its surroundings, does not show a reduction in size of the intergranular lanes. In fact the lanes increase in size by about 31\% in the time between the minimum B-field and the peak. However, a merger between two MBPs is observed to occur during the amplification of the B-field. This occurs between 153~s to 306~s after detection. Two small intensity enhancements can be seen between 1$''$ to 1.5$''$ in $x$ and 2$''$  in $y$ in the intensity images in the initial frame (Figure~\ref{Fig9}). These gradually move closer to each other before coalescing at about 272~s. The magnetic field rises to 1375~G at this point as the two have merged into a single entity. This is similar to the scenario we found in our observations. This case is perhaps clearer than our observations, as the merger event in our observations occurred between two MBPs in a group of several MBPs. Here, there are only two in close proximity so it is clearer that the merger event is likely responsible for the B-field amplification.

We note that actual merger events are not as frequent in the simulations, possibly due to the fact that the FOV is smaller. Also, it is not possible to fully resolve intergranular lane flux at the same resolution in our observations due to the combination of a lower spatial resolution and polarimetric sensitivity.

\subsection{Amplification due to vorticity}
\label{vort}

The MURaM simulations also show that magnetic field can be amplified when the MBP moves within a region of higher vorticity. We do not find evidence for this process in our observations.

\begin{figure}[h!]
\makebox[\linewidth]{
   \includegraphics[width=0.9\linewidth]{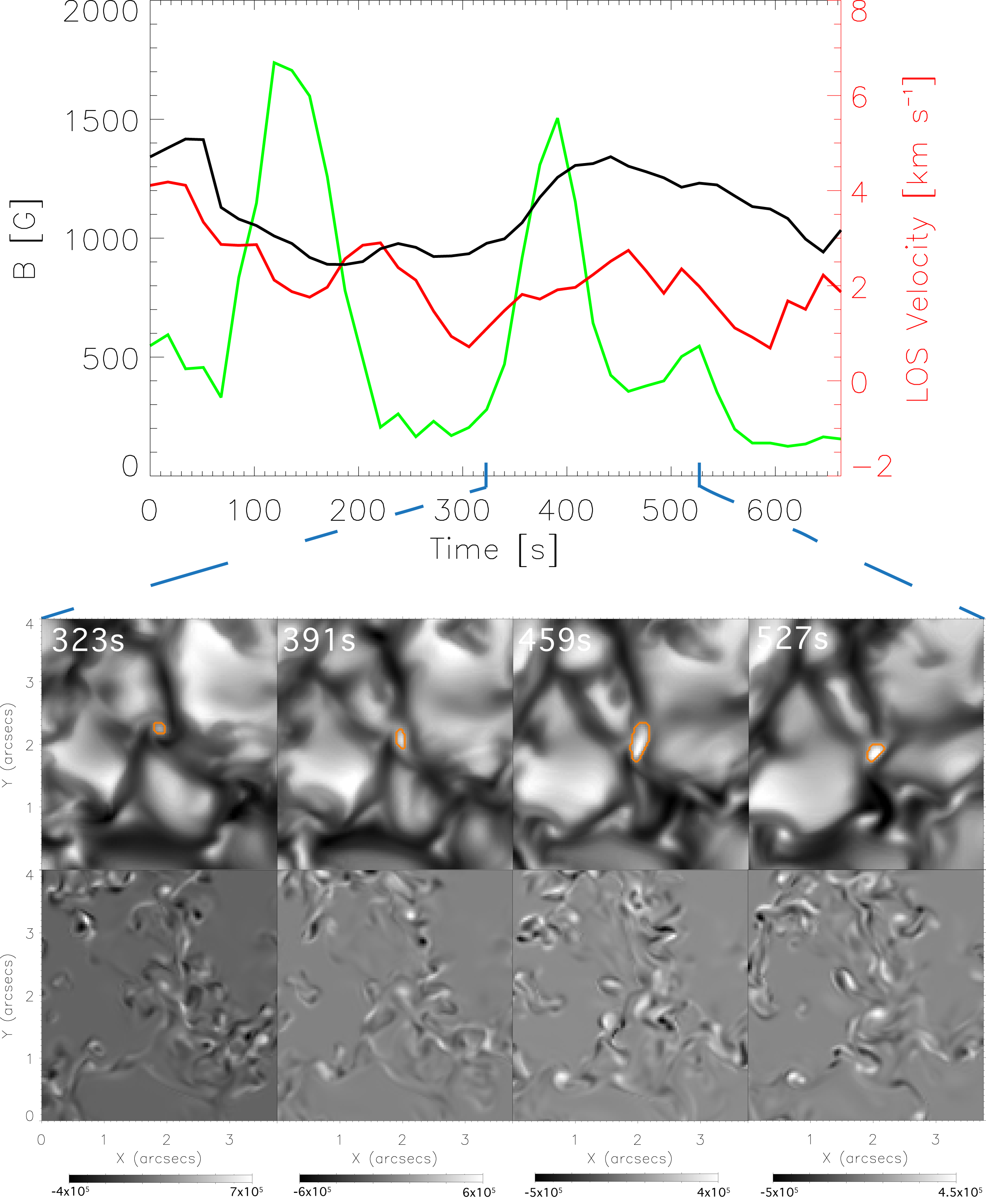}
	}
     \caption{Example from our MURaM simulations of vorticity leading to magnetic field amplification in an MBP. The black line shows the magnetic field in the vertical direction while the red line indicates the LOS velocity. The green line shows the vorticity within the MBP during its evolution, established from the velocity components of the simulations. The panels below show the evolution of the MBP in the intensity images (top) and the evolution of the vorticity for the corresponding intensity images, for the time period indicated by the blue dashed line. Orange contours in the intensity images show the MBP, as tracked by our algorithm, used to make the evolution plots above. The MBP appears to rotate prior to a magnetic field amplification and intensity enhancement, which corresponds to the increase in the vorticity in the vicinity of the MBP.
}
     \label{Fig10}
\end{figure}

Figure~\ref{Fig10} shows an example of vorticity within the simulations resulting in magnetic field amplifications. In this case, the MBP (which has a lifetime of 680~s) has a relatively low B-field, 918~G, prior to a peak of 1359~G 170~s later. In this time the LOS velocity decreases below 1~km\,s$^{-1}$, with a peak of 2.5~km\,s$^{-1}$ 153~s before the B-field begins to rise. Visual inspection of the MBP shows that it experienced rotational motion just prior to the increase in B-field and throughout the magnetic field amplification phase. An examination of the baroclinic vorticity (green line in Figure~\ref{Fig10}, obtained from the velocity components of the simulated domain) related to the MBP across these frames shows an increases by a factor of 14 just prior to the amplification of the B-field, relaxing afterwards, resulting in the MBP B-field reducing again. It takes the barcolinic vorticity 102~s to increase from the minimum to its peak value prior to the B-field peak. The B-field peaks 51~s after the peak in vorticity, with the vorticity relaxing again in the same time frame. The intensity and vorticity maps (lower panel images in Figure~\ref{Fig10}) show that vorticity within the region of the MBP increases prior to the magnetic field amplification and intensity enhancement. The actual size of the MBP does not vary significantly during this process, varying by about 6\% between 408~s after detection and 527~s after initial detection. 

Work by \citet{Shelyag2011} found evidence for two types of vortex in MBPs, namely a baroclinic vortex from the hydrodynamics term and magnetic vortices due to magnetic tension. We observe both here in this MBP. The authors state that magnetic vortices have a direct link to rotational motions within the MBP, which we observe with this MBP and that magnetic vortices may have a link to swirl motions observed higher up in the atmosphere. \citet{Shelyag2011} showed that magnetic vorticity due to magnetic tension is responsible for most vorticity observed in the upper photosphere, while the baroclinic vorticity contributions have increasing importance with increasing geometrical depth. 

As well as an increase in baroclinic vorticity, we also find an increase in the magnetic vorticity in the MBP over a similar time period, prior to a peak in B-field. The magnetic vorticity for this MBP (not shown in Figure~\ref{Fig10}) increases by a factor of 3 from 255~s after initial detection, peaking at 323~s after initial detection. The time between the peak in magnetic vorticity and the peak in B-field is 119~s. Magnetic vorticity in the MBP decreases to a minimum again about 459~s after initial detection of the MBP, 136~s after the peak in magnetic vorticity and 51~s after the peak in B-field. Given the height that these images are taken from in the simulated domain, it is possible that both the baroclinic and magnetic vorticity within the MBP contribute to the amplification in the magnetic field to some extent. It is somewhat difficult to disentangle the relative contributions of both to the amplification of the field. However, it is interesting that both values drop to a minimum near the peak of the B-field, which then results in the B-field gradually declining over time.

Another interesting facet of this MBP (though not shown here) takes place around the first peak in baroclinic vorticity within the MBP (i.e. from about 85~s to 170~s after initial detection). In this time frame the vorticity within the MBP rises sharply with the MBP splitting in two from a larger feature to two smaller entities. There is a subtle internal rotation within the MBP as it splits, as well as a slight curvature in motion of the MBP after the split with a peak in vorticity in the vicinity of the MBP. The B-field of the MBP dips in this time frame from 1085~G at the point before the split levelling out at about 885~G after the peak in vorticity within the MBP. This drop in B-field is clearly due to the split in the feature. This reiterates the complex evolutionary characteristics imposed on MBPs due to the plasma motions near the MBP. As the MBP gets stretched out in this instance, it is weakened at the boundary between the two newly forming features, which will allow a greater flow at the boundary. These different flows across the MBP likely lead to the vortical motion within the MBP.

This is the only concrete case that we have found in our simulations of the effects of vorticity on the evolution of B-field properties in MBPs. We do not find evidence for such amplification in our observations, however, we point this unique case out to stress that it may be possible that vortical motions result in field amplifications for MBPs. This will be the focus of a future work.


\section{Concluding remarks}
\label{conc}
In this work, a high spatial and temporal resolution dataset was acquired in full Stokes spectropolarimetry mode at the Swedish Solar Telescope of a quiet Sun region at disc centre. MBPs within the dataset were tracked to determine their evolutionary properties. The physical properties, such as LOS velocity and LOS magnetic field strength, were inferred with respect to atmospheric height by inverting the Stokes parameters. This was performed using two publicly available inversion codes, namely NICOLE \citep{SocasNavarro2015} and SIR \citep{CoboIniesta1992,BellotRubio2003}. The inversion codes provided consistent results.

Previous work on this dataset showed that MBPs in our FOV had a bimodal distribution in terms of magnetic field strength \citep{Keys2019}. Within the work presented here it was observed that MBPs could effectively transition between the `weak' and `strong' groups multiple times within their lifetimes, with relatively short transitions times ($\sim$33\,--\,99~s). Three possible processes for the rapid amplification in the field were observed and described. 

These amplification processes include: the well-known `convective collapse' process \citep{Spruit79}, amplification due to granular compression and amplification due to the merging of two or more neighbouring MBPs. Convective collapse was the most frequently occurring process, with the LOS velocity increasing prior to the B-field amplification. The MBP shrinks to achieve equipartition, balancing the pressure inside and outside the flux tube, resulting in an amplification of the B-field. With granular compression, a similar shrinkage in the MBP's size (as its shape becomes more elliptical) results in amplification of the magnetic field as neighbouring granules expand and compress it. This process could have implications for wave studies, as the compression of the flux tube is similar to the drivers often observed to excite sausage modes. The final process identified in the observations was due to merging MBPs. Upon the merging of MBPs, the magnetic field is observed to increase, possibly due to additional flux residing within the `new' MBP and/or the new feature becoming slightly more compressed on merging. It was observed that all three processes could occur at multiple times during the MBP lifetimes, leading to multiple peaks in B-field. Future work employing better temporal resolution in the spectropolarimetric data will be utilised to determine more accurately the timescales over which these amplifications occur.

Furthermore, we employed MURaM simulations \citep{Vogler2005} to ascertain whether similar events could be found in MBPs within a simulated domain. Indeed, we detect the same three processes in the simulations too. Also, we see evidence for vortical motions leading to magnetic field amplification in MBPs within the simulations, though we do not see similar cases in the observations. Whether such a case exists in the observations will be examined in future work.

Within this work, we see that complex behaviour appears to be intrinsic to MBP evolution, which is an important factor to consider in fields such as wave propagation in MBPs, as well as flux emergence and evolution in the lower solar atmosphere. This behaviour can potentially complicate the interpretation of observations and should be considered when analysing MBPs. Future missions, such as DKIST and EST, may elucidate further on these evolutionary properties of MBPs, by improving the spatial resolution and polarimetric sensitivity at which they are observed.


\begin{acknowledgements}
{All authors are grateful to the anonymous referee for suggestions to improve the manuscript. P.H.K. is grateful to the Leverhulme Trust for the award of an Early Career Fellowship. P.H.K would like to thank Drs. Luis Bellot Rubio, David Kuridze and Hector 
Socas-Navarro for assistance in setting up the inversion codes properly and the HAO/NCAR/NSO sponsored ASP workshop on spectropolarimetry for general advice on interpreting inversion results. R.L.H. would like to thank the Department of Education and Learning of Northern Ireland for the award of a Ph.D. studentship. A.R, M.M., and V.J.M.H. acknowledge support from the Science and Technology Facilities Council (STFC) under grant No. ST/P000304/1.
D.B.J. wishes to thank the UK STFC for the award of an Ernest Rutherford Fellowship alongside a dedicated Research Grant. D.B.J. also wishes to thank Invest NI 
and Randox Laboratories Ltd. for the award of a Research \& Development Grant (059RDEN-1) that allowed this work to be undertaken. S.J. acknowledges support from the European Research Council (ERC) under the European Union's Horizon 2020 research and innovation program (grant agreement No. 682462). V.M.J.H. was supported by the Research Council of Norway, project number 250810. Both V.M.J.H. and S.J. were also supported by the Research Council of Norway through its Centres of Excellence scheme, project number 262622.
Observations were acquired within the 
SolarNet Project Transnational Access Scheme. SolarNet is a project supported by the EU-FP7 under Grant Agreement 312495. The Swedish 1-m Solar Telescope is operated on the island of La Palma by the Institute for Solar Physics of Stockholm University in the Spanish Observatorio del Roque de los Muchachos of the Instituto de Astrof{\'{i}}sica de Canarias. The Institute for Solar Physics is supported by a grant for research infrastructures of national importance from the Swedish Research Council (registration number 2017-00625). We acknowledge support from the STFC Consolidated Grant to Queen's University Belfast.}
\end{acknowledgements}


\end{document}